\documentclass[cits]{PoS}
\usepackage{amsmath,amssymb}
\usepackage{graphicx}
\newcommand{\nn}{\nonumber}
\newcommand{\Vq}{{\cal V}_{q}}
\newcommand{\Vp}{V^{+}}
\newcommand{\Vm}{V^{-}}
\newcommand{\wt}{{\omega_\tau}}
\newcommand{\Feff}[1]{{\cal F}_\mathrm{#1}}
\title{Chiral and deconfinement transitions\\
in strong coupling lattice QCD}

\ShortTitle{Chiral and deconfinement transitions in SC-LQCD}

\author{\speaker{Kohtaroh Miura}\\
        INFN Laboratori Nazionali di Frascati, 
	I-00044, Frascati (RM), Italy\\
        E-mail: \email{Kohtaroh.Miura@lnf.infn.it}}

\author{Takashi Z. Nakano\\
	Department of Physics, Faculty of Science, Kyoto University, 
	Kyoto 606-8502, Japan\\
        Yukawa Institute for Theoretical Physics, 
	Kyoto University, Kyoto 606-8502, Japan}

\author{Akira Ohnishi\\
	Yukawa Institute for Theoretical Physics, 
        Kyoto University, Kyoto 606-8502, Japan}

\author{Noboru Kawamoto\\
	Deparment of Physics, Faculty of Science, Hokkaido University, 
	Sapporo 060-0810, Hokkaido, Japan}

\abstract{
We investigate the QCD phase diagram
based on the strong coupling expansion of the lattice QCD
with one species of the staggered fermions
at finite temperature ($T$) and chemical potential ($\mu$).
We analytically derive an effective potential
including both chiral and deconfinement ($Z_{N_c}$)
dynamics with finite coupling effects
in mean-field approximations.
We focus on Polyakov loop properties
in whole $T-\mu$ plane, and study
relations between the chiral and deconfinement
crossovers.
At a fixed large $\mu$,
sequencial rapid variations
of the Polyakov loop are observed
with increasing $T$.
It is natural to interprete them as
the ``chiral induced'' and ``$Z_{N_c}$ induced''
deconfinements crossovers.}
\FullConference{The XXVIII International Symposium on Lattice Field Theory, Lattice2010\\
		June 14-19, 2010\\
		Villasimius, Italy}
\begin{document}
\section{Introduction}
The strongly interacting non-Abelian gauge theory
is a central key-concept in the high energy physics.
One of the most important properties
is the asymptotic freedom,
which indicates a phase transition (or crossover)
from hadron phase to a deconfined matter
-Quark-Gluon Plasma (QGP)-
in Quantum Chromodynamics (QCD)
when temperature ($T$) and/or 
quark chemical potential ($\mu$)
exceed a typical energy scale of QCD ($\Lambda_{\mathrm{QCD}}$).
To investigate the QGP and this phase transition
is a physics goal
in the LHC-ALICE experiment, and indispensable
to understand the astrophysical systems,
such as the early universe and the neutron stars.

The phase transition (or crossover)
from the hadron to the QGP
can be described in terms of
two kinds of (approximate) symmetries, 
the SU($N_f$) $\times$ SU($N_f$) chiral symmetry
and the center $Z_{N_c}$ symmetries
($N_c$ and $N_f$ represents the number of colors and flavors).
The finite value of the Polyakov loop
indicates a sort of deconfinement 
in the sense that an excitation energy
of heavy quark is finite.
The finite value of the chiral condensate
at vanishing current quark mass
indicates the hadron phase in the sense that
hadrons are massive (the pion is massless)
and relevant degrees of freedom.
The Polyakov loop and the chiral condensate
breaks the $Z_{N_c}$ and the chiral symmetries, 
respectively,
and are typical observables in the lattice QCD.
The lattice QCD based investigation
for the phase structure of their symmetries
{\em i.e.} the {\em QCD Phase Diagram}
in $T-\mu$ plane is one of the most important subjects
in the high energy physics.

The Lattice QCD Monte-Carlo simulations (LQCD-MC)
indicate that chiral and deconfinement ($Z_{N_c}$)
crossovers simultaneously happen around,
$T_c=\simeq 160-190~\mathrm{MeV}$ at $\mu=0$~\cite{Muller:2006ee}.
The coincidence of ``$Z_{N_c}$ and chiral''
is the non-trivial observation, and the relation
between their crossovers is
a modern interest on the QCD phase diagram.
The important sampling method
in LQCD-MCs suffers from the
sign problem of the quark determinant with finite $\mu$,
and the LQCD-MC does not work well at finite $\mu$.
It is required to shed light on
the large $\mu$ region of the QCD phase diagram
beyond the sign problem.

In this proceedings,
we investigate QCD phase diagram 
in the whole range of $T-\mu$ plane
by using
the {\em Strong Coupling Expansion in Lattice QCD (SC-LQCD)}~\cite{Montvay}
with one species of (unrooted) staggered fermion.
The analytic investigation based on SC-LQCD 
can be informative for finite $\mu$,
and there are many recent developments
\cite{NishiFuku,Azcoiti:2003eb,Kawamoto:2005mq,deForcrand:2009dh,NLO,Nakano:2009bf,Nakano:2010bg}.
We take account of
next-to-leading order (NLO, ${\cal O}(1/g^2)$) terms
in the strong coupling expansion.
In addition, 
we consider the leading pure gluonic contributions
for Polyakov loops.
We concentrate on the leading order
of the large dimensional ($1/d$)
expansion~\cite{KlubergStern:1982bs} for simplicity.
In these setups,
we focus on the relation between
the $U_{\chi}(1)$ chiral and $Z_{N_c}$ deconfinement dynamics
in the phase diagram.

\section{Effective Potential}\label{sec:Feff}
We start from the lattice QCD partition function
with one species of staggered fermion ($\chi$)
with a quark mass ($m_0$).
Gluons are represented
by link variables ($U_{\nu,x}$), 
\begin{align}
Z= \int\mathcal{D}[\chi,\bar{\chi},U_{\nu}]
\exp\biggl[
-\sum_{\nu,x}
\frac{\eta_{\nu,x}\bar{\chi}_x U_{\nu,x} \chi_{x+\hat{\nu}}
-\eta_{\nu,x}^{-1}(h.c.)}{2}
-\sum_x m_0(\bar{\chi}\chi)_x
+\sum_P\frac{U_P+U_P^{\dagger}}{g^2}
\biggr]
\label{eq:Z}\ .
\end{align}
where,
$U_{P=\mu\nu,x}=\mathrm{tr}_c[U_{\mu,x}U_{\nu,x+\hat{\mu}}U^\dagger_{\mu,x+\hat{\nu}}U^\dagger_{\nu,x}]$,
and $(\eta_{0,x},\eta_{j,x})=(e^{\mu},(-1)^{x_0+\cdots +x_{j-1}})$.
By using a $\gamma_5$-related 
factor $\epsilon_x=(-1)^{x_0+\cdots+x_{d}}$,
a staggered chiral transformation is defined as
$\chi_x\to e^{i\theta\epsilon_x}\chi_x$~\cite{Montvay}.
The lattice action
is invariant under this $U_{\chi}(1)$
chiral transformation
in the chiral limit $m_0\to 0$.
In the following,
we use the lattice unit $a=1$,
and we investigate
the color SU($N_c=3$) case in $3+1$ dimensions ($d=3$).

\begin{figure}[ht]
\includegraphics[width=7.8cm]{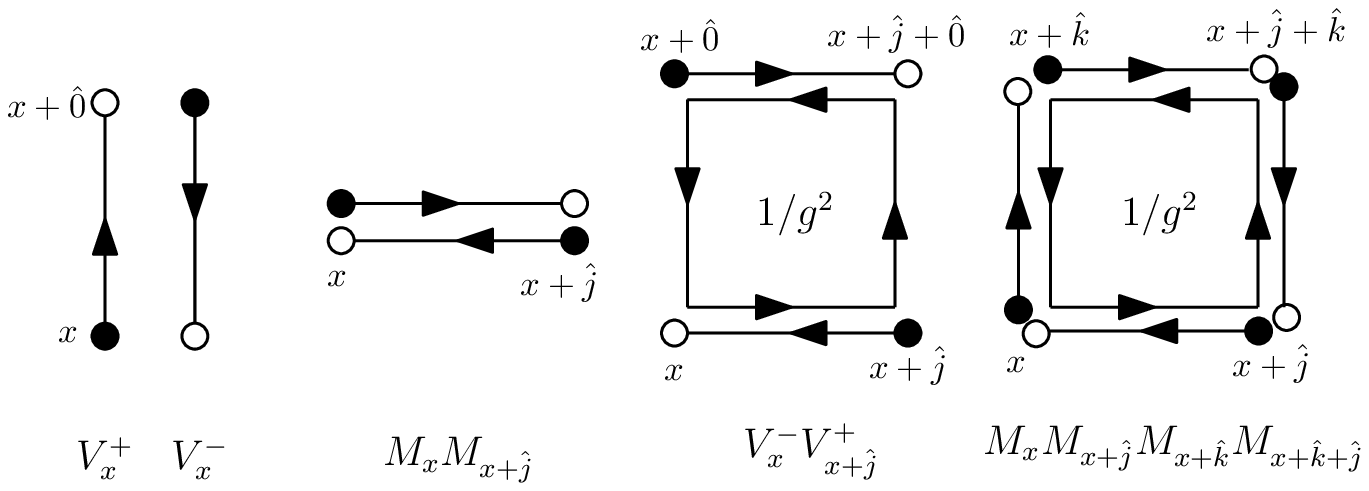}
\includegraphics[width=7.0cm]{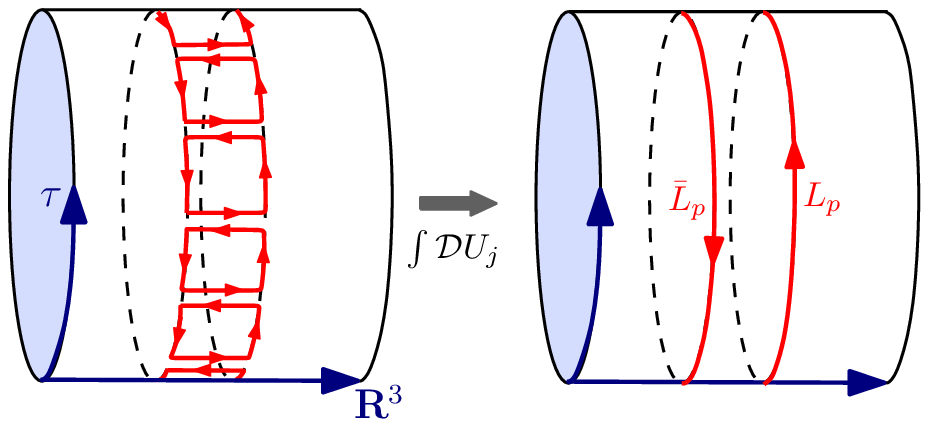}
\caption{Left:~The strong coupling limit 
and $1/g^2$ effects to appear 
within the leading order of $1/d$ expansion.
Open circles, filled circles, and arrows 
show $\chi$, $\bar{\chi}$, and
$U_\nu$, respectively.
Right:~Schematic figure of
the pure gluonic Polyakov loop
in the strong coupling expansion.}
\label{Fig:diagrams}
\end{figure}

We perform strong coupling expansions,
and take account of the next-to-leading order (NLO) effects.
To manipulate temperature effects,
we firstly evaluate the spatial gluon degrees of freedom
by utilizing the group integral formula,
$\int dU_{jx}~{U_{jx}}_a^b{U_{jx}^{\dagger}}_c^d=\delta_a^d\delta_c^b/N_c$,
which leads to sums of color singlet 
(hadronic) composites.
The spatial link integral leaves 
an isotropic sum over spatial directions,
and the energy per bond is assumed to be proportional to $1/d$
in order to keep the action finite at large spatial dimension $d$
\cite{KlubergStern:1982bs}.
In this normalization scheme,
the quark field scales as $d^{-1/4}$.
We concentrate on
the leading order $1/d$ terms, which
correspond to the minimum quark number
configurations for a given plaquette geometry.
Within these approximations,
we obtain the following hadronic composites
and effective couplings,
\begin{align}
&(\beta_{\tau},~\beta_s)
=\Bigl(\frac{d}{N_c^2g^2},~\frac{d(d-1)}{8N_c^4g^2}\Bigr)
=\Bigl(\frac{d}{2N_c^3}\beta,~\frac{d(d-1)}{16N_c^5}\beta\Bigr)
\ ,\\
&(M_x,~\Vp_x(\mu),~\Vm_x(\mu))
=(\bar{\chi}_x\chi_x,~
e^{\mu}\bar{\chi}_xU_{0,x}\chi_{x+\hat{0}},~
e^{-\mu}\bar{\chi}_{x+\hat{0}}U_{0,x}^{\dagger}\chi_{x})\ ,
\end{align}
where the lattice coupling $\beta=2N_c/g^2$ is the unique
parameter in the current theory.
The left panel of Fig.~\ref{Fig:diagrams}
provides a schematic description of composites.
In addition to them,
we extract the leading pure gluonic effect
to Polyakov loops
($L_{p,\mathbf{x}}=\prod_{\tau}U_{0,x}$,
the right hand side of Fig.~\ref{Fig:diagrams}),
which is necessary to investigate
a deconfinement dynamics.
These setups give us the simplest framework
to investigate both $U_{\chi}(1)$ chiral and
$Z_{N_c}$ deconfinement transitions (or crossovers) in
the SC-LQCD with finite coupling effects.
Recent our study indicates that
the next-to-next-to leading order effects
with quarks has just a small correction
to the phase diagram,
while there is discrepancy between SC-LQCD
and the Lattice QCD Monte-Carlo simulations (LQCD-MC)
for the $\beta=2N_c/g^2$ evolutions of the critical
temperature at zero chemical potential, 
($T_{c,\mu=0}(\beta)$)~\cite{Nakano:2009bf}.
Due to the pure gluonic Polyakov loops,
the $T_{c,\mu=0}(\beta)$ is found to become
closer to the lattice Monte-Carlo (LQCD-MC)
simulations around $\beta=4.0$
at $\mu=0$~\cite{Nakano:2010bg}.
We anticipate that
the pure gluonic Polyakov loops
could have an essential contribution
at ``finite $\mu$'', which will be investigated here.

To integrate 
the quark degrees of freedom ($\chi,\bar{\chi}$),
we introduce
four kinds of mean fields~\cite{NLO},
which are summarized in Table~\ref{Tab:aux}.
The mean field $\sigma$ represents
the chiral condensate, which leads
to the dynamical mass shift,
$m_0\to m_q=m_0+(d/(2N_c))\sigma$
in the strong coupling limit
It is remarkable that the NLO effects are expressed as
modifications of
quark mass and chemical potential,
$(m_q,\mu)\to(\tilde{m}_q,\tilde{\mu})$ with
the quark wave function renormalization factor,
$\sqrt{Z_{+}Z_{-}}$, and those are governed by
the unique lattice parameter $\beta$~\cite{NLO},
\begin{align}
\tilde{m}_{q}
=\frac{1}{\sqrt{Z_{+}Z_{-}}}
\biggl(
m_0+\Bigl(\frac{d}{2N_c}+2\beta_s\varphi_s\Bigr)\sigma
\biggr)\ ,\quad
\tilde\mu
=\mu-\log\sqrt{\frac{Z_+}{Z_-}}
\ ,\label{eq:Renorm}
\end{align}
where,
$Z_{\pm}=1+\beta\bigl(\varphi_\tau\pm\omega_\tau\bigr)$.

This NLO formulation which is invented
in our previous work~\cite{NLO}
allows us to evaluate
the remnant degrees of freedom,
quarks ($\chi,\bar{\chi}$)
and the temporal link variable $U_0$,
in the same way to
the strong coupling limit case~\cite{DKS,IK_GO}.
It is convenient to take a static and diagonalized gauge
(called the Polyakov gauge)
for temporal link variables with respect for the periodicity~\cite{DKS},
$U_{0,\mathbf{x}}=\mathrm{diag}\bigl\{e^{i\theta^1_{\mathbf{x}}/N_\tau},\cdots,e^{i\theta^{N_c}_{\mathbf{x}}/N_\tau}\bigr\}$.
In this gauge, the Polyakov loop reduces to a simple expression,
$L_{p,\mathbf{x}}\to \sum_{a=1}^{N_c}e^{i\theta^a_{\mathbf{x}}}\equiv N_cl_{p,\mathbf{x}}$,
and the Haar measure is also expressed 
in terms of the Polyakov loops,
\begin{align}
\int dU_{0,\mathbf{x}}
&\to 
\int_0^1 d[l_{p,\mathbf{x}},\bar{l}_{p,\mathbf{x}}]~
27\mathcal{M}_{\mathrm{Haar}}(l_{p,\mathbf{x}},\bar{l}_{p,\mathbf{x}})\ ,\\
\mathcal{M}_{\mathrm{Haar}}(l_{p,\mathbf{x}},\bar{l}_{p,\mathbf{x}})
&=
1-6\bigl(\bar{l}_{p,\mathbf{x}}l_{p,\mathbf{x}}\bigr)
-3\bigl(\bar{l}_{p,\mathbf{x}}l_{p,\mathbf{x}}\bigr)^2
+4\bigl(l_{p,\mathbf{x}}^{N_c}+\bar{l}_{p,\mathbf{x}}^{N_c}\bigr)
\ .\label{Eq:LpHaar}
\end{align}
The Haar measure effects are responsible 
for the deconfinement dynamics.
The quark determinant includes
the quark propagation
wrapping the temporal direction with $U_0$,
and leads to the quark driven Polyakov loops,
which is characterized by,
\begin{align}
\mathcal{D}_{q}(T,\mu)
&\equiv
1+N_c\bigl(l_{p,\mathbf{x}}e^{-(E_q-\tilde{\mu})/T}
+\bar{l}_{p,\mathbf{x}}e^{-2(E_q-\tilde{\mu})/T}\bigr)
+e^{-N_c(E_q-\tilde{\mu})/T}\ ,\label{eq:Dq}\\
\mathcal{D}_{\bar{q}}(T,\mu)
&\equiv
1+N_c\bigl(\bar{l}_{p,\mathbf{x}}e^{-(E_q+\tilde{\mu})/T}
+l_{p,\mathbf{x}}e^{-2(E_q+\tilde{\mu})/T}\bigr)
+e^{-N_c(E_q+\tilde{\mu})/T}\ .\label{eq:Dqb}
\end{align}
Here,
$E_q(\tilde{m}_{q}(\sigma))=\sinh^{-1}\bigl[\tilde{m}_{q}(\sigma)\bigr]$
corresponds to the quark excitation energy.
The Polyakov loop $l_p$ couples 
to a Boltzmann factor $e^{-(E_q-\tilde{\mu})/T}$,
and determines quarks thermal excitations.
Note that the Boltzmann factors is a function of
the chiral condensate $\sigma$,
therefore
the chiral and deconfinement dynamics
couples to each other 
via Eq.~(\ref{eq:Dq}) and (\ref{eq:Dqb}).
This effect has been investigated
in the strong coupling limit case ~\cite{IK_GO},
and now we have the NLO corrections via
Eq.~(\ref{eq:Renorm}).
Instead of integrating out the temporal link variables,
we replace the Polyakov loop with 
its constant mean-field value, 
$(l_{p,\mathbf{x}},\bar{l}_{p,\mathbf{x}})\to (l_{p},\bar{l}_{p})$.
Finally, we obtain the effective potential
as a function of the auxiliary fields
$\Phi=(\sigma,\varphi_{\tau,s},\omega_{\tau},l_p,\bar{l}_p)$,
temperature $T$, and 
quark chemical potential $\mu$,
\begin{align}
\Feff{eff}(\Phi;T,\mu)
=&
\Bigl(
\frac{d}{4N_c}+\beta_s\varphi_s
\Bigr)\sigma^2
+\frac{\beta_s\varphi_s^2}{2}
+\frac{\beta_{\tau}}{2}
\bigl(
\varphi_{\tau}^2-\omega_{\tau}^2
\bigr)-N_c\log \sqrt{Z_{+}Z_-}\nn\\
&-N_cE_q-T\bigl(
\log\mathcal{D}_{q}(T,\mu)
+\log\mathcal{D}_{\bar{q}}(T,\mu)
\bigr)\nn\\
&-2dTN_c^2\biggl(\frac{1}{g^2N_c}\biggr)^{1/T}\bar{l}_{p}l_{p}
-T\log\mathcal{M}_{\mathrm{Haar}}(l_p,\bar{l}_p)
\ .\label{eq:Feff}
\end{align}
The equilibrium is determined
by imposing stationary conditions
on the effective potential, 
$\partial{\Feff{eff}}/\partial\Phi=0$,
which leads to the relations summarized
in the third column of Table \ref{Tab:aux}.

\begin{table}[ht]
\caption{The auxiliary fields and their stationary values.
Here,
$\varphi_0=N_c-\sqrt{Z_{+}Z_-}\tilde{m}_q+\beta_{\tau}\omega_{\tau}^2$.
}\label{Tab:aux}
\begin{center}
\begin{tabular}{c|c|c}
\hline
Aux. Fields&Mean Fields &Stationary Values\\
\hline
$\sigma$        &$\langle -M\rangle$
                &$-\partial\Vq/\partial(\sqrt{Z_+Z_-}\tilde{m}_{q})$\\
$\varphi_s$     &$\langle MM\rangle$
                &$\sigma^2$\\
$\varphi_\tau$  &$-\langle (\Vp-\Vm)/2\rangle$
                &$2\varphi_0/(1+\sqrt{1+4\beta_\tau\varphi_0})$\\
$\wt$           &$-\langle (\Vp+\Vm)/2\rangle$
                &$-\partial\Vq/\partial\tilde{\mu}=\rho_q$\\
\hline
\end{tabular}
\end{center}
\end{table}

\section{Results}\label{sec:res}
\begin{figure}[ht]
\begin{center}
\includegraphics[width=7.0cm]{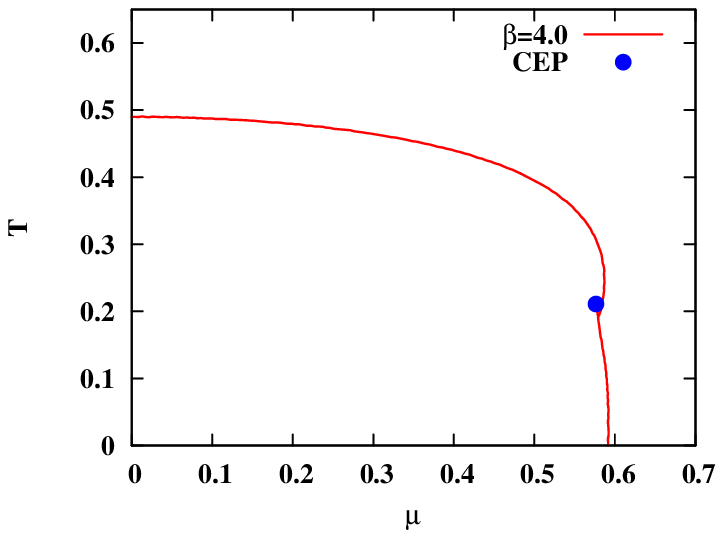}
\includegraphics[width=7.8cm]{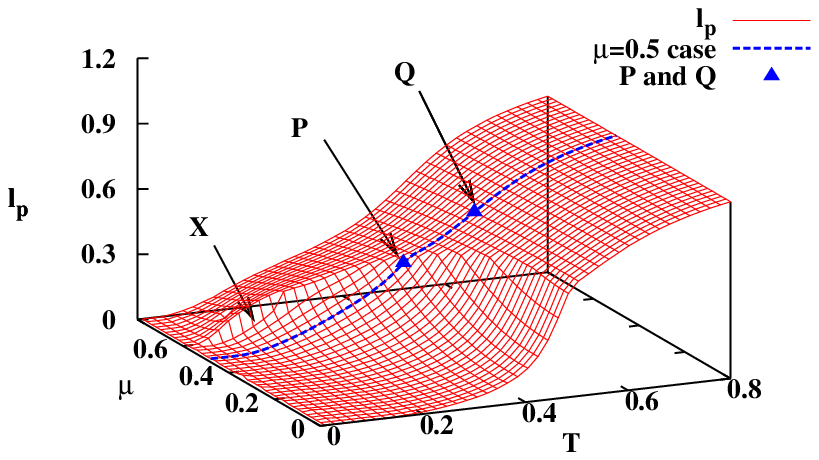}
\caption{Left:~
The phase diagram in the chiral limit at $\beta=4.0$.
Right:~ The equilibrium value of the Polyakov loop
as a function of $T$ and $\mu$ in the chiral limit
at $\beta=4.0$.}
\label{Fig:PD}
\end{center}
\end{figure}

We will concentrate on the $\beta=4.0$ results
in the followings.
In the left panel of Fig.~\ref{Fig:PD},
we show the phase diagram in the chiral limit.
In the low $T$ and large $\mu$ region,
we find the critical end point (CEP),
and observe a partially restored (PCR) matter
around the CEP. The details of PCR have been explained
in our previous paper~\cite{NLO}.
In this proceedings, we concentrate on the
relation between the chiral and deconfinement crossovers.
In the right panel of the Fig.~\ref{Fig:PD},
the equilibrium value of the Polyakov loop $l_p$ is
depicted as a function of $Ta$ and $\mu a$
in the same condition as the left panel.
We find the rapid variation of $l_p$
around the chiral transition line.
Thus the deconfinement and chiral dynamics
are strongly correlated except zero temperature cases.

\begin{figure}[ht]
\begin{center}
\includegraphics[width=7.4cm]{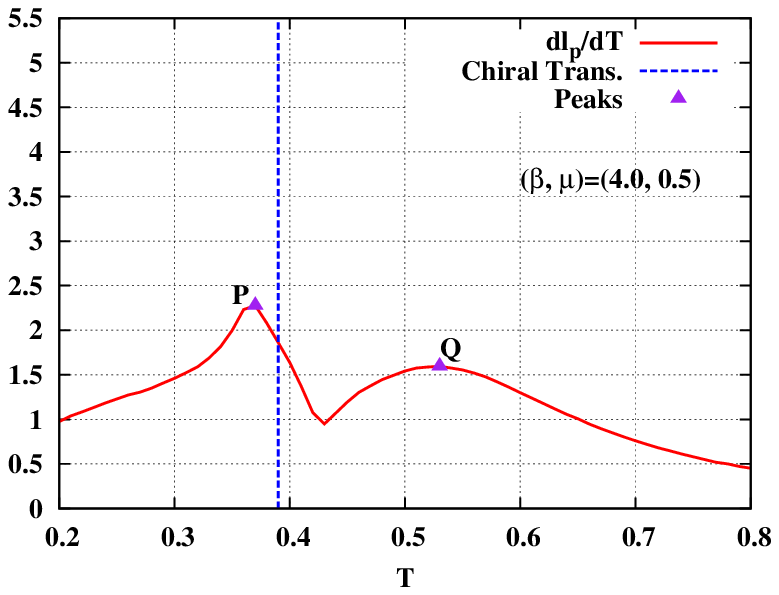}
\includegraphics[width=7.4cm]{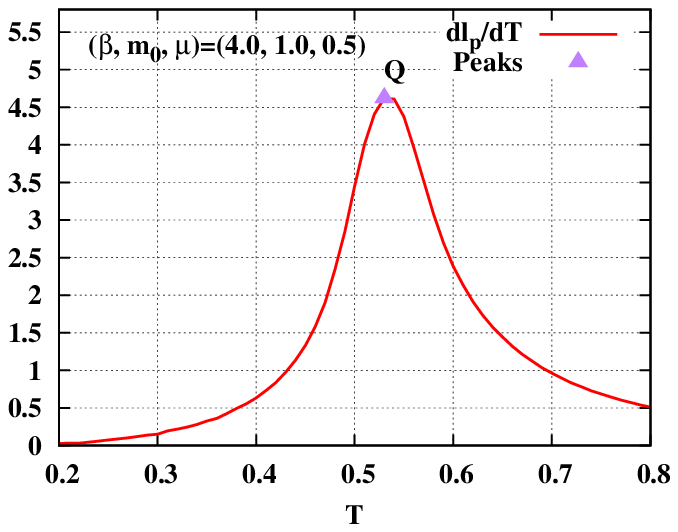}
\caption{The derivative of the Polyakov loop
in terms of $T$ at $\mu=0.5$ in the lattice unit.
The left and right panels show 
the chiral limit and heavy mass ($m_0a=1.0$)
cases, respectively.
In the left panel,
the vertical dashed (blue) line represents
chiral transition temperature.}
\label{Fig:DLp}
\end{center}
\end{figure}

Let us go along with the $\mu a=0.5$ (dashed blue) line
in the right panel of the Fig.~\ref{Fig:PD}.
The derivative of the Polyakov loop in terms of $T$
on this line is shown in the left panel of
Fig.~\ref{Fig:DLp}.
In the chiral limit ($m_0\to 0$, left panel),
we find two peaks ``P'' and ``Q'',
and the former corresponds to the steepest
point on the $\mu a=0.5$ line 
in the right panel of Fig.~\ref{Fig:PD}
(Here, ``P'' and ``Q'' in the Fig.~\ref{Fig:DLp}
correspond to those in the Fig.~\ref{Fig:PD}).
Near the first peak ``P'',
the Polyakov loop is force to change its value
to minimize the effective potential
due to the drastic variation of the chiral condensate.
In this meaning, the peak ``P'' can be
interpreted 
as the {\em Chiral Induced Deconfinement Crossover}.
When the current quark mass $m_0$ becomes large,
the $Z_{N_c}$ center symmetry becomes dominant.
The right panel in the Fig.~\ref{Fig:DLp}
corresponds to such a situation ($m_0a=1.0$).
We find that the second peak ``Q'' grows up,
and the ``P'' disappears.
This indicates that the second peak
is the {\em $Z_{N_c}$ Induced Deconfinement Crossover}.

It is interesting that
the $Z_{N_c}$ nature still survives in the chiral limit,
and its $\mu$ dependence is almost negligible.
Hence the chiral and $Z_{N_c}$ dynamics
start to separate with increasing $\mu$.
This is the mechanism of the two sequential
deconfinement crossovers.
We note that the coincidence of 
the chiral and $Z_{N_c}$ at ``$\mu=0$''
is still mysterious, and to be investigated in future.

\section{Summary}\label{sec:sum}
We have investigated the chiral and deconfinement crossovers
at finite temperature ($T$) and chemical potential ($\mu$)
based on the strong coupling expansion in the lattice QCD
with one species of the staggered fermion.
We have considered the next-to-leading order (NLO) effects
in the strong coupling expansion with the leading pure gluonic
contributions to the Polyakov loops.
We have concentrate on the leading order of
the $1/d$ expansion, and derived the analytic expression
of the effective potential in the framework of the mean-field approximation.
We have replaced the Polyakov loop
with its constant mean fields value
without integrating the temporal link variable,
and the deconfinement dynamics is introduced via
the Haar measure of temporal link integral.

At a fixed large $\mu$,
we have observed sequential 
two deconfinement crossovers {\em i.e.}
the rapid variations
of the Polyakov loops as $T$ increases.
One of them results from the strong correlation
between the chiral and deconfinement dynamics.
In other word, the Polyakov loop
is force to change its value
to minimize the effective potential
due to the drastic variation of the chiral condensate.
In this meaning, we have found
the {\em Chiral Induced Deconfinement Crossover}.
When the current quark mass $m_0$ becomes large,
this crossover disappears, and
the another crossover grows up.
Thus we have also observed
the {\em $Z_{N_c}$ Induced Deconfinement Crossover}.
In the chiral limit,
the former becomes stronger,
and the latter becomes weaker.
The point is that the latter still exists
in the chiral limit.
Hence two crossovers co-exists there.

There are several studies to be performed in future.
Firstly, it has been known that
the temporal link integral can be analytically performed,
and then the fluctuation effects appear
as the combination of the modified Bessel function
and the difference of Polyakov loops
in the effective potential~\cite{Nakano:2010bg}.
It should be confirmed that the obtained results
in this work is independent of approximation schemes.
Secondly,
the chiral and Polyakov loop susceptibilities
would be informative to investigate
the relation between the chiral and deconfinement dynamics.
Thirdly, the exact evaluation in each order of
the strong coupling expansion is required beyond
the $1/d$ expansion and the mean-field approximation.
This could be achieved by extending the
Monomer-Dimer-Polymer formulation \cite{deForcrand:2009dh,MDP}
to include the NLO effects.
And finally, it would be interesting to introduce
the advanced formulation of the pure gluonic
effects for the Polyakov loops which is recently
investigated in Refs.~\cite{Langelage}.

\section*{Acknowledgments}
We thank to Prof. Maria Paola Lombardo for
the fruitful discussions.
This work was supported in part
by Grants-in-Aid for Scientific Research from MEXT and JSPS
(Nos. 22-3314),
the Yukawa International Program for Quark-hadron Sciences (YIPQS),
and by Grants-in-Aid for the global COE program
``The Next Generation of Physics, Spun from Universality and Emergence''
from MEXT.


\begin{thebibliography}{99}

\bibitem{Muller:2006ee}
For a recent review, see
  B.~M\"uller and J.~L.~Nagle,
  Ann.\ Rev.\ Nucl.\ Part.\ Sci.\ {\bf 56}, (2006) 93.


\bibitem{Montvay}
The review of the pioneering works for 
the strong coupling expansion is found in the text book,
  I.~Montvay and G.~M\"unster,  
  ``Quantum Fields on a Lattice,''
  Cambridge~University~Press,~1994.



\bibitem{NishiFuku}
  Y.~Nishida, K.~Fukushima and T.~Hatsuda,
  Phys.\ Rept.\  {\bf 398} (2004), 281;
  K.~Fukushima,
  Prog.\ Theor.\ Phys.\ Suppl.\  {\bf 153} (2004), 204;
  Y.~Nishida,
  Phys.\ Rev.\  D {\bf 69} (2004), 094501.

\bibitem{Azcoiti:2003eb}
  V.~Azcoiti, G.~Di Carlo, A.~Galante and V.~Laliena,
  J. High Energy Phys. {\bf 09} (2003), 014.

\bibitem{Kawamoto:2005mq}
  N.~Kawamoto, K.~Miura, A.~Ohnishi and T.~Ohnuma,
  Phys.\ Rev.\  D {\bf 75} (2007), 014502.

\bibitem{deForcrand:2009dh}
  P.~de Forcrand and M.~Fromm,
  Phys.\ Rev.\ Lett.\  {\bf 104} (2010) 112005.



\bibitem{NLO}
  K.~Miura, T.~Z.~Nakano, A.~Ohnishi and N.~Kawamoto,
  Phys.\ Rev.\  D {\bf 80} (2009) 074034;
  K.~Miura, T. Z~Nakano and A.~Ohnishi,
  Prog.\ Theor.\ Phys.\  {\bf 122} (2009), 1045.


\bibitem{Nakano:2009bf}
  T.~Z.~Nakano, K.~Miura and A.~Ohnishi,
  Prog.\ Theor.\ Phys.\  {\bf 123} (2010) 825.


\bibitem{Nakano:2010bg}
  T.~Z.~Nakano, K.~Miura and A.~Ohnishi,
  arXiv:1009.1518 [hep-lat];
  also appear in this proceedings.


\bibitem{KlubergStern:1982bs}
  H.~Kluberg-Stern, A.~Morel and B.~Petersson,
  Nucl.\ Phys.\  B {\bf 215} (1983), 527.
\bibitem{DKS}
%
  P.~H.~Damgaard, N.~Kawamoto and K.~Shigemoto,
  Nucl.\ Phys.\  B {\bf 264} (1986), 1.


\bibitem{IK_GO}
  E.~M.~Ilgenfritz and J.~Kripfganz,
  Z.\ Phys.\  C {\bf 29} (1985), 79;
  A.~Gocksch and M.~Ogilvie,
  Phys.\ Rev.\  D {\bf 31} (1985), 877;
  K.~Fukushima,
  Phys.\ Rev.\  D {\bf 68} (2003), 045004.


\bibitem{MDP}
  F.~Karsch and K.~H.~Mutter,
  Nucl.\ Phys.\  B {\bf 313}, (1989), 541.




\bibitem{Langelage}
  J.~Langelage and O.~Philipsen,
  JHEP {\bf 1004} (2010) 055;
  JHEP {\bf 1001} (2010) 089;
  J.~Langelage, G.~Munster and O.~Philipsen,
  JHEP {\bf 0807} (2008) 036.

\end{thebibliography}
\end{document}